\documentstyle[prb,amssymb,preprint,aps,epsf]{revtex}
\begin{document}
\bibliographystyle {prsty}
\draft
\title{
Resonance in  One--Dimensional Fermi--Edge Singularity.
}
\author{Yuval Oreg and Alexander M. Finkel'stein\cite{Lan}}
\address{
Department of Condensed Matter Physics, The Weizmann Institute of Science,
Rehovot 76100, Israel}
\date{January 96}
\maketitle

\makeatletter
\global\@specialpagefalse
\def\@oddhead{}
\let\@evenhead\@oddhead
\makeatother

\begin{abstract}
\label{abstract}

The problem of the Fermi--edge singularity in a one--dimensional
Tomonaga--Luttinger liquid is reconsidered.  The backward scattering
of the conduction band electrons on the impurity--like hole in the
valence band is analyzed by mapping the problem onto a Coulomb gas
theory. For the case when the electron--electron interaction is
repulsive the obtained exponent of the one--dimensional Fermi--edge
singularity appears to be different from the exponent found in the
previous studies.  It is shown that the infrared physics of the
Fermi--edge singularity in the presence of backward scattering
and electron--electron repulsion resembles the physics of
the Kondo problem.

\end{abstract}


    \section{Introduction}

    \label{intro}

Large optical singularities in the absorption and emission spectra
have been observed by Calleja et. al.  \cite{XR:Calleja91} in
semiconductor quantum wire structures in the extreme quantum limit
when only the lowest one--dimensional (1--d) subband is occupied by
electrons. The pronounced features in the optical spectra have been
interpreted as strong Fermi--edge singularities (FES) of the 1--d
electron gas. The FES arises (e.g. for absorption), because the
optical process is accompanied by a multiple scattering of the
conducting electrons on a hole in the valence band, which is created
at the absorption of the electro--magnetic wave and acts as an
impurity--like center.  In the case of 1--d geometry the scattering of
the conducting electrons by the hole is limited to scattering in the
forward or backward directions.  Another special feature is related to
the electron--electron interaction in 1--d. In contrast to higher
dimensions, the 1--d electron gas can not be treated as an ordinary
Fermi liquid even when the electron--electron interaction is small.
The experiment \cite{XR:Calleja91} initiated a number of papers
discussing the FES in 1--d metallic systems for the forward scattering
\cite{XR:Ogawa92,XR:DLee92} and the backward one
\cite{XR:Gogolin93,XR:Prokofev94,XR:Kane94,XR:Affleck94} in the
presence of the electron--electron interaction.

In the present paper the effect of the backward scattering on the FES
in 1--d will be reconsidered. The exponent of the 1--d FES obtained
here appears to be different from the one that was found in Ref.'s
\cite{XR:Gogolin93,XR:Prokofev94,XR:Kane94,XR:Affleck94}.  It is shown
that the infrared physics of the Fermi--edge singularity in the
presence of backward scattering together with electron--electron
repulsion resembles the physics of the Kondo problem. This aspect of
the problem was missed in the preceding studies.

FES or Mahan singularity is a power law singularity in the
electro--magnetic wave absorption (or emission) coefficient when the
frequency is close to the Fermi energy \cite{XR:Mahan67b}.  That is in
contrast to a naive expectation that the absorption coefficient is
zero bellow the Fermi energy and is proportional to the density of
states of the conduction band above the Fermi energy. Since in common
metals the FES is observed in the $X$--ray range, it is also called
sometimes the $X$--ray singularity.  The singularity arises because
the absorption is accompanied by the infrared catastrophe phenomena.
When the electro--magnetic wave is absorbed, an electron from a deep
level is excited to the conduction band and, correspondingly, a hole
is created deep in the valence band. The abrupt appearance of the
scattering potential of the deep hole leads to excitation of an
infinite number of low energy electron--hole pairs in the conduction
band. Besides, the electron excited from the deep level to the
conduction band scatters multiply on the hole. The latter process,
which includes an exchange with other electrons of the conduction
band, leads to enhancement of the absorption coefficient.  The
physical reason of this enhancement is the increase of the effective
density of states near the deep hole due to its attraction of
electrons. On the other hand the creation of the electron--hole pairs
in the conduction band reflects the fact that the final state of the
electrons in the presence of the deep hole potential is orthogonal to
the initial state.  This process is called the orthogonality
catastrophe of Anderson \cite{XR:Anderson67} and it leads to a
reduction of the absorption coefficient.  The value of the exponent of
the FES is determined by these two competing effects.

The effects related to the physics of the FES can be found in various
situation where a sudden change of a scattering potential happens.
The well known example is the Kondo problem describing scattering of
conduction electrons on a magnetic impurity. When the impurity spin is
flipped the impurity potential on which the conduction electrons are
scattered is abruptly altered. Therefore, the Kondo problem can be
treated \cite{KE&AMM:Yuval70} as a response of the electrons on a
sequence of sudden changes of the local scattering potential.  The
problem can be mapped on a classical Coulomb gas theory of alternating
positive and negative charges (spin flips) with logarithmic
interaction.  In the present paper it will be shown that the FES
problem in 1--d can also be analyzed by mapping it on a Coulomb gas
theory, similarly to the Kondo problem.

To calculate the exponent of the FES it is needed to sum a series of
infrared--divergent logarithmic terms.  A corresponding theory was
developed for three dimensional metals (3--d) by Nozi\'eres and
collaborators in a series of papers
\cite{XR:Nozieres69a,XR:Nozieres69b,XR:Nozieres69c}.  The recoil of
the hole, due to its finite mass, together with Auger processes lead
to a natural cutoff of the infrared singularities.  It is usually
assumed, for simplicity, at the discussion of the FES that these
effects can be neglected. The other common assumption is that the
potential created by the deep hole is a short range one.  That is
because the electron screening develops on short scales, while at
studies of infrared phenomena large distances are essential.  Since in
3--d metals the Fermi liquid parameters are determined by the short
scale physics, it is also generally assumed that one can use the Fermi
liquid description for the conduction electrons at the consideration
of the FES.

At studies of the FES in 1--d the role of these common assumptions may
change.  It has been noticed \cite{XR:Ogawa92} that in 1--d systems
the recoil of the deep hole does not suppress the FES, at least for
the forward scattering.  Here we will ignore the recoil effects
completely as well as the Auger processes. In this paper we will also
not discuss the problem of screening of the potential of the deep
hole, despite that it may be essential for the FES in
modulation--doped quantum wires. Since we are mainly interested now in
the effects of the $2k_F$--backward scattering, it will be assumed
that the potential of the deep hole is local. The most intriguing
aspect of FES in 1--d is related to the role of the electron--electron
interaction.  In some sense the Fermi surface is weakened in 1--d as a
result of the interaction of electrons. In the case of the
Tomonaga--Luttinger model the jump in the occupation numbers of
electrons at the Fermi energy is replaced by a singularity, and there
are no single electron quasi--particles
\cite{EE1D:Mattis65,EE1D:Dzyaloshinskii74,EE1D:Luther74}.  Since the
FES is a Fermi surface effect, the investigation of the role of the
electron--electron interaction in this problem is of clear interest.

The effective way to investigate the infrared properties of the 1--d
electron systems is the bosonization technique. In fact, the important
step in the developing of this technique was made just in connection
with the problem of the FES. Schotte and Schotte extended the approach
of Ref. \cite{EE1D:Mattis65} and developed a very compact and
economical description of the FES for $s$-wave scattering in 3--d
\cite{XR:Schotte69a}.  In the present paper the consideration of the
electrons interaction will be confined to the Tomonaga--Luttinger
model where it is assumed that the electrons do not scatter backward
in the course of the electron--electron interactions. Since
Tomonaga--Luttinger model has a simple solution in the bosonization
technique, it is natural to apply that technique for finding the FES
in 1--d in the presence of interactions between the electrons.

In the case when at the scattering on the potential of the hole in the
valence band the electrons are scattered only in the forward
direction, the FES can be found \cite{XR:Ogawa92} following the Schotte and
Schotte method.  For completeness of the
presentation and with the purpose of introducing notations we have
reconsidered this case in Sec. \ref{se:fs}. When the backward
scattering on the hole potential is included, it is shown in
Sec. \ref{se:bs} that the problem can be mapped on a gas of charged
particles, similarly to the Coulomb gas mapping in the Kondo problem
\cite{KE&AMM:Yuval70,KE&AMM:Anderson70a,KE&AMM:Schotte70}.  The role
of the spin flip in the Kondo problem acts now the change of the
electron motion from left to right and visa versa at the backward
scattering on the hole.  In comparison with the Kondo problem the
Coulomb gas describing the 1--d FES is different in two aspects.  In
the $S=\frac 12$ Kondo problem successive flips should follow
alternatively, while in the case of FES there is no limitations on the
order of left--right and right--left scattering. 
(In the end of Sec. \ref{se:ri} a certain relationship of the non
alternating Coulomb gas with the two--channel Kondo problem is
outlined.) The other difference
is due to the creation of the additional electron in the conduction
band at the absorption of the electro--magnetic wave. This gives rise
to two additional charges located at the ends of the Coulomb gas
system.  In the case of repulsion of electrons the gas of the charged
particles is in the hot plasma phase.  Therefore a new scale --- the
screening radius of the plasma --- is generated in this problem as a
combined effect of the backward scattering and the interaction.  The
shape of the absorption line for frequencies close enough to the
Fermi--edge is controlled by this scale.  The reduction of the 1--d
FES problem to the classical gas theory yields a value of the FES
exponent in the vicinity of the Fermi edge, which is different from
the one obtained in preceding studies
\cite{XR:Gogolin93,XR:Prokofev94,XR:Kane94,XR:Affleck94}.

In section \ref{se:NL} the FES is analyzed using fields, which where
first introduced in Ref. \cite{XR:Kane94}.  In that representation the
scattering is simplified, but at a price that the electron--electron
interaction is described by a non--local theory of self--dual fields.
A special treatment is needed in case of such a theory.  The
renormalization group analysis of the problem using these variables is
elaborated in Sec. \ref{se:rg}. In addition, an iteration procedure is
developed in Sec. \ref{se:qtsp} for the analysis of the FES problem.
This procedure is useful for determining the screening length of the
Coulomb gas in the hot plasma phase.

 To understand the obtained exponent of the FES, in Sec. \ref{se:ri}
the result is interpreted in terms of the phase shift theory of
Nozi\'eres and De Dominicis \cite{XR:Nozieres69c}.  It is argued that
the discrepancy between the value of the FES exponent obtained in the
present paper and the result of Ref.'s
\cite{XR:Gogolin93,XR:Prokofev94,XR:Kane94,XR:Affleck94} is because in
these papers the asymptotic regime of the problem is described by a
weak link junction.  However, two weakly connected wires represent a
system with strong depletion of the density of electrons at the defect
center. It is shown in the end of Sec. \ref{se:ri} that the use of
such system as the asymptotic limit description of a weak impurity
scatterer, as it was suggested in the theory of Ref.
\cite{EE1D:Kane92,EE1D:Kane92a},
 is in contradiction with the Friedel sum rule. In
fact, it is concluded here that the infrared asymptotic regime of the
FES resembles to the Kondo resonance.

    \section{Coulomb gas}

    \label{se:phiphidual}

In this section we show that in 1-d systems the correlation function $F(t)$
determining (after the Fourier transform) the Fermi edge singularity
can be presented as a product of two terms. The first one is related to the
forward scattering of the conduction electrons by the hole, which is created
in the valence band at the absorption of the external electro--magnetic
wave. This term can be calculate directly, while the second one describing
the backward scattering by the hole, is analyzed by mapping onto a Coulomb
gas. This approach allows us to find the behavior of $F(t)$ at
asymptotically large time. For compactness we confine the consideration
bellow to a spinless case, and the spin degrees of freedom of the conduction
electrons are included in the appendix (\ref{app}).

\subsection{The forward scattering}

\label{se:fs} 

The Hamiltonian describing  the 1-d electron liquid is given by

\begin{equation}
\label{eq:ffh}
H_0=
 iv_F \int dx \psi^{\dagger}_R(x) \frac{\partial}{\partial x} \psi_R(x)
-iv_F \int dx \psi^{\dagger}_L(x) \frac{\partial}{\partial x} \psi_L(x)+
  \frac 12 V   \int dx \left( \rho_R(x) + \rho_L(x) \right) ^2,
\end{equation}
where $\psi_R  (x)  \mbox  {  and  } \psi^\dagger_R (x) $ are the field
operators of fermions that propagate to the right with wave vectors $\approx
+k_F$, and $\psi_L (x) \mbox { and } \psi^\dagger_L (x) $ are the field
operators of left propagating fermions with wave vectors $\approx -k_F$; $\;
\rho_{L\left( R \right)} (x)= \psi^\dagger_{L(R}(x) \psi_{L \left( R \right) } (x) $ are the
electron density operators; the spectrum of the electrons is linearized near
the Fermi points and $v_F$ is the Fermi velocity; $V$ describes the
density--density interaction with momentum transfers much smaller than $k_F$.
Hamiltonian (\ref{eq:ffh}) corresponds to the Tomonaga--Luttinger model,
which describes the 1-d electron liquid when the backward scattering
amplitude of the electron-electron interaction may be ignored.  Eq.
(\ref{eq:ffh}) is a fixed point Hamiltonian for a broad class of 1-d systems.

 After the absorption of the external electro--magnetic wave a hole is
created in the valence band together with an additional electron in the
conduction band. It will be assumed hereafter that the position of the
hole is fixed at $x=0$. The scattering of the conduction band electrons by
this hole is given by

\begin{equation}
\label{eq:fih}
    H_{sc} = U(k=0)  \left( \psi_R^\dagger(0) \psi_R  (0) + \psi^\dagger_L  (0)
\psi_L (0) \right) + U(2 k_F) \psi^\dagger_R (0) \psi_L(0) +  U^*(2 k_F) 
\psi^\dagger_L (0) \psi_R (0) ,
\end{equation}
    where $U(k)$ are the Fourier transform amplitudes of the hole potential $U(x)$,  and
$  \psi_{R(L)}  (0)  =  \psi_{R(L)}  (x=0)  $. Thus, $U(k=0)$ is the forward
scattering amplitude and $U(2k_F) = -|U(2k_F)| e^{i \varphi_u}$ is the backward one.

    In order  to find the electro--magnetic wave   absorption line shape
$I(\omega)$ one needs to calculate the Fourier transform of the correlation
function \cite{XR:Nozieres69c,XR:Schotte69a}
\begin{equation}
\label{eq:F} F(t)=\left< e^{iH_0t} \psi(0) e^{-iH_ft} \psi^\dagger(0) \right>,
\end{equation}
    where $\psi^\dagger (0)$ refers to an electron which is created in
the conduction band at the absorption, and $H_f=H_0+H_{sc}$ is the
Hamiltonian describing the electron liquid after the creation of the
hole.  It has been shown in the seminal paper of Schotte and Schotte
\cite{XR:Schotte69a} that bosonic representation of the fermion
operators in Eq. (\ref{eq:F}) gives an illuminating approach for
understanding the FES in the case of $s-$wave scattering by the hole.
It is natural to apply this approach for the analysis of the FES in
1-d conductors.

It is well known \cite{EE1D:Mattis65} that the bosonization technique (for
review see \cite{EE1D:Solyom79,RFS:Fradkin91}) allows to reduce the Tomonaga--Luttinger
Hamiltonian to a quadratic form in terms of operators of bosonic fields
$\phi_0$ and $\tilde \phi_0$:

\begin{mathletters}
\label{eq:dpd} 
\begin{eqnarray}
\phi_0(x)&=&\frac{-i}{\sqrt{4\pi}} \frac{2\pi}{L} \sum_p \frac{\exp\left({-\eta |p|/2-ipx}\right)}{p}
\left[ \rho_R(p)+\rho_L(p) \right],\\
\tilde\phi_0(x)&=&\frac{-i}{\sqrt{4\pi}} \frac{2\pi}{L} \sum_p \frac{\exp\left({-\eta |p|/2-ipx}\right)}{p}
\left[ \rho_R(p)-\rho_L(p) \right], 
\end{eqnarray}
\end{mathletters}
    where $\eta^{-1}$ is  an ultraviolet cutoff,  which is of the  order of
the conduction band width, and $L$ is the system length. The fields
$\phi_0$ and its dual partner $\tilde \phi_0$ are conjugate variables, i.e.,
\begin{equation}
\label{eq:cr}
\left[\frac{d\phi_0(x)}{dx},\tilde \phi_0(y) \right]=i\bbox{\delta}(x-y).
\end{equation}
 After rescaling the operators
 \begin{equation}
\label{eq:rphi}
\phi = \frac {\sqrt {4\pi}} {\beta} \phi_0 , \;
\tilde \phi = \frac {\beta} {\sqrt{4\pi}} \tilde \phi_0,
\end{equation} 
the bosonized representation of Hamiltonian (\ref{eq:ffh}) becomes
\begin{equation}
\label{eq:bfh}
H^B_0=\frac{1}{2} v_F \alpha \int dx \left(
    \left(  \frac {d\phi}{dx} \right)^2
+   \left( \frac {d \tilde \phi}{dx} \right)^2
 \right),
\end{equation}
where
\begin{equation}
\label{eq:abc}
\alpha=\frac{4\pi}{\beta^2},\; \beta^2=4\pi \sqrt {\frac {1-\gamma}
 {1+\gamma} } \mbox { and } \gamma=\frac{V}{\left( 2\pi v_F+V \right)}.
\end{equation}
 In the
 bosonization technique \cite{EE1D:Solyom79,RFS:Fradkin91} the bosonic
 representations of the operators $\psi_R$ and $\psi_L$ are given as
\begin{mathletters}
\label{eq:fermion}
\begin{eqnarray}
\psi^B_R(x)&=&
\frac {e^{ ik_Fx}} {\sqrt{2\pi\eta}} \exp{\left[- \frac{i}{2} \left(
\frac {4\pi} {\beta} \tilde\phi + \beta \phi \right) \right] } \\
\psi^B_L(x)&=&
i\frac {e^{-ik_Fx}} {\sqrt{2\pi\eta}} \exp{\left[-  \frac{i}{2} \left(
\frac {4\pi} {\beta} \tilde\phi - \beta \phi \right) \right]}.
\end{eqnarray}
\end{mathletters}
Then, the scattering by the hole may be written in
terms of the $\phi$-field as

\begin{equation}
\label{eq:bih}
 H^B_{sc}= \frac{1}{2} v_F  \left(
    \frac{\beta \delta_+}{ \pi} \left. \frac{d\phi}{dx} \right|_{x=0}
  -\frac{2\delta_-}{\pi\eta} \cos(\beta \phi(0) + \varphi_u) \right),
\end{equation}
where the dimensionless amplitudes 
$\delta_+=\frac {U(k=0)}{v_F}  \mbox{ and } \delta_- = \left| 
\frac {U(2k_F)}{v_F} \right|$ are introduced and
 $\varphi_u =\arg  \left( -\frac{U(2k_F)}{v_F} \right)$.

In Eq. (\ref{eq:bih}) the term related to the forward scattering is linear
in $\phi$. Therefore, a canonical transformation that shifts at the
point $x=0$ the operator $\phi^\prime(x)= \frac{d \phi}{dx}$ can be
exploited to exclude the forward scattering.
 Since
 \begin{equation}
\label{eq:tp}
 e^{ ia \tilde\phi(0)} \phi^\prime(x) e^{ -ia \tilde\phi(0)} =
\phi^\prime(x)+a\bbox{\delta}(x),
\end{equation}
 the transformed Hamiltonian
 $ \tilde H_f^B=e^{ia \tilde\phi(0)} \left( H_0^B+H_{sc}^B \right) e^{
 -ia \tilde\phi(0)}$ will not contain the forward scattering term, if
$a=-\frac{1}{2\pi}\frac{\beta}{\alpha} \delta_+$.
In the absence of the backward scattering the correlation function
$F(t)$, after carrying out the unitary transformation ${\hat{\cal U}}(0)=\exp \left(
\frac{i}{2\pi} \frac{\beta}{\alpha} \delta_+\tilde \phi (0) \right)$,
becomes:

\begin{equation}
\label{eq:fb}
F(t)_{forward}=\left<
 \exp \left( iH_0^Bt \right)
 \psi(0)
\hat{\cal U}(0)
 \exp \left( -i  H_0^B t \right)
\hat{\cal U}(0)^{-1}
\left( \psi(0) \right)^\dagger
  \right>.
\end{equation}
The calculation of $F(t)_{forward}$ can be reduced to a Gaussian--like
integral if the bosonic representation (\ref{eq:fermion}) for the
operators $\psi(0)$, $\psi^\dagger(0)$ is applied. Let us choose the 
component $\psi_R(0)$ for the operator $\psi(0)$; the contribution of
the other component is  equal.  Then, the use of the bosonic
representation gives:
\begin{equation}
\label{eq:fb1}
F(t)_{forward}=
\left<
 e^{-i\frac 12 \beta \phi(x=0,t)}
 e^{-i \frac 12 \tilde \beta \tilde \phi(x=0,t)} 
 e^{i \frac 12 \tilde \beta \tilde \phi(x=0,0)}
 e^{i\frac 12 \beta \phi(x=0,0)}
\right>,
\end{equation}
where $\tilde \beta = \frac{4\pi}{\beta}- \frac{\delta_+ \beta}{\pi
\alpha}$, and the operators $\phi(t)$ and $\tilde \phi(t)$ depend on time as
$\hat O(x,t) =e^{iH_0^Bt} \hat O (x,t=0) e^{-iH_0^Bt}$. Now, the standard
application of the Baker-Hausdorff formula $e^A e^B = e^{A+B} e^{ \frac {1}
{2} [A,B] } $ together with the Gaussian averaging $\left< e^{iA} \right> =e^{-
\frac {1} {2} \left< A^2 \right> } $ yield
\begin{equation}
\label{eq:fb2}
F(t)_{forward}=
e^{-\frac{1}{4} \beta^2 {\cal G}_{phi}(t)
 -\frac{1}{4} \tilde \beta^2 {\cal G}_{\tilde \phi}(t)},
\end{equation}
where ${\cal G}_{\phi}(t)=\left< \phi(0,t)\phi(0,0)-\phi(0,0)^2 \right>$ is the Green
function  of the $\phi$--operators, and ${\cal G}_{\tilde \phi} $ is defined similarly.
It has also been assumed here that
$ 
\left< \tilde \phi (t) \phi (0) \right> = \left< \phi (t) \tilde \phi
(0)\right> = 0$,
what agrees with Eqs. (\ref{eq:dpd}).
From Eqs. (\ref{eq:dpd}) and (\ref{eq:bfh}) it can be obtained for
asymptotically large time
\begin{equation}
\label{eq:Green}
{\cal G}_{\phi}(t)={\cal G}_{\tilde \phi}(t)= \frac{1}{2\pi} \log \left( 1+i\frac{\alpha v_F t}{\eta}
\right).
\end{equation} 
Finally one gets:
\begin{equation}
F_{forward} \sim \left( \frac{1}{t} \right)^{1-\alpha_{forward}}  \;\;\;\;\;\;
  \alpha_{forward}= 1 - \frac{1}{8 \pi}
\left( \tilde \beta^2+\beta^2 \right)
\end{equation}
 and correspondingly  the absorption line shape $I(\omega) \sim 
 \left( \delta \omega \right)^{ -\alpha_{forward} }$, where $\delta \omega = \omega -
\omega_{threshold}$ and 
\begin{equation}
\label{eq:Fw}
\alpha_{forward} = \left( 1 - \frac{1}{\sqrt{1-\gamma^2}} \right) +
\frac{\delta_+}{\pi}
 \sqrt {\frac{1-\gamma}{1+\gamma}} - \frac{\delta_+^2}{2 \pi^2} \left( \sqrt
{ \frac{1-\gamma}{1+\gamma} } \right) ^3. \end{equation} Here the first term reflects the
Luttinger--liquid behavior of the 1--d electrons in the presence of an
electron--electron interaction \cite{EE1D:Dzyaloshinskii74,EE1D:Luther74}.
Note that contrary to the noninteracting case the last term in Eq.
(\ref{eq:Fw}), which corresponds to the Anderson orthogonality catastrophe,
is not the half of the square of the second term.

 The above procedure was elaborated  by Schotte and Schotte
\cite{XR:Schotte69a} for consideration of the FES in 3-dimensional case for
$s$-wave scattering. For the 1--d case the function $F_{forward}(t)$,
including the spin degrees of freedom, has been obtained recently in Ref.
\cite{XR:Ogawa92} (see bellow Eq. (\ref{eq:betas}) for comparison).

 \subsection{The backward scattering}
\label{se:bs}
In the presence of the backward scattering 
 the correlation function $F(t)$  will be studied  in the interaction
representation with respect to $\tilde H^B_{b-sc}$:
\begin{equation}
\label{eq:fint}
F(t) = \left< \exp \left( i H_0^B t \right) \psi (0) \hat{\cal U} (0) \exp \left( - i H_0^B t \right)
{\cal T} \exp \left( - i \int_0^t \tilde H_{b-sc}^B (t^\prime ) d t^\prime \right) \hat{\cal U} (0)^{-1} \left(
\psi (0) \right) ^\dagger \right>,
\end{equation} 
where ${\cal T}$ is the time ordering symbol and  the backward scattering term is
\begin{equation}
\label{eq:bsh}
\tilde H_{b-sc}^B=-\frac{v_F \delta_-}{\eta \pi} \hat {\cal U}(0)^{-1}
\cos \left( \beta \phi(0)+\varphi_u \right) \hat {\cal U}(0).
\end{equation}
 (The action of the electron--electron
interaction, unlike the backward scattering term, is not limited to the interval
$[0,t]$. Therefore, electron transitions  not only within this time
interval contribute to the FES. Fortunately, in the case of the
Tomonaga--Luttinger model the renormalization of the fields $\phi$ and
$\tilde \phi$ by Eq. (\ref{eq:rphi}) takes that complications into
consideration, and Eq. (\ref{eq:fint}) holds.)
  When an expansion of the ${\cal T}$-exponential in $\tilde H_{b-sc}^B(t^\prime)$ is
performed, the result can be written as a sum of all possible time--ordered
products of the exponentials of the operators $\pm i \left[ \beta
\phi(x=0,t^\prime) + \varphi_u \right] $. The averaging procedure of each term
in this sum can be easily performed using repeatedly the Baker--Hausdorff
formula. This is a standard way of a reduction of the quantum problem of
such kind to the physics of classical charged gases; see for example Refs.
\cite{KE&AMM:Anderson70a,KE&AMM:Schotte70,EE1D:Chui75,EE1D:Legget87}.

Now a subtle point related to the application of the Baker--Hausdorff
formula should be commented.
 Eq. (\ref{eq:cr}) determines the
commutation relations of $\phi$ and $ \tilde \phi$ only up to a constant: 
\begin{equation}
\label{eq:icr}
\left[ \phi(x), \tilde \phi(0) \right]= i\int^x
 \bbox{\delta}(x^\prime)dx^\prime
= i\frac 12 \mbox{ sign } (x) +C.
\end{equation}
This constant may be fixed by the requirement that the representation
(\ref{eq:fermion}) should ensure the fermion commutation relations of
the operators $\psi^B_L$ and $\psi^B_R$. That requirement leads to
$C=\pm \frac{i}{2}$.  In Eq.~(\ref{eq:bih}) the cosine of the backward
scattering term has been written with $C= - \frac{i}{2}$.  Now, using
the fact that in $F(t)$ the creation and annihilation operators of the
right and left moving electrons  appear in equal numbers for non
vanishing terms, one can check in a formal way that $C$ does not
influence $F(t)$.  For that reason it will be assumed hereafter that
\begin{equation}
\label{eq:0cr}
\left[ \phi(0), \tilde \phi (0) \right] =0,
\end{equation}
what is in correspondence with  a naive treatment of this commutator  with
 representation~(\ref{eq:dpd}). 
Consequently,   when the Gaussian averaging is performed in the calculation 
of the correlation function $F(t)$ it may be assumed that 
\begin{equation}
\label{eq:zero_correlation} 
\left< \tilde \phi (t) \phi (0) \right> = \left< \phi (t) \tilde \phi
(0)\right> = 0,
\end{equation} 
Usually the correlation function of $\phi$ and $\tilde \phi$
operators controls the relative phase of certain operators when they
exchange their coordinates, because

\begin{equation}
\label{eq:AB}
\left< \tilde \phi (x,i\tau) \phi (y,i\tau^\prime) \right> =\frac{1}{2\pi}\arctan \left(
\frac{x-y}{\tau-\tau^\prime} \right).
\end{equation}
In some problems the angle appearing in this equation has been
interpreted as the Aharonov--Bohm phase
\cite{RFS:Jose77,XR:Kadanov78}. The peculiarity of the problem at
hand is in the fact that the operators in the correlation
function~$F(t)$ are at a single point $x=0$. Consequently, in
the~$(x,t)$ plane all the operators stay along a straight line, and
for that reason the Aharonov--Bohm phases can be chosen to be zero.

 Finally this procedure yields
\begin{equation}
\label{eq:f1}
F(t)= e^{-\frac{1}{4} {\tilde \beta}^2 {\cal G}_{\tilde \phi}(t)} Z_{\phi} (t),
\end{equation}
where
\begin{eqnarray}
\label{eq:fphi}
&{\displaystyle Z_{\phi}(t)= \sum_{n=0}^\infty \left( \frac{ \delta_-}{2\pi
\alpha} \right) ^n
  \sum_{q_t,q_0,q_{\mu}} {}^{^{\displaystyle \!\!\!\!\prime}}
 i^n  e^{\left( i \varphi_u \sum_\mu q_\mu \right)} 
\int_0^t \frac{\alpha v_F}{\eta} dt_n \ldots \int_0^{t_3} \frac{\alpha
v_F}{\eta} dt_2
\int_0^{t_2} \frac{\alpha v_F}{\eta} dt_1}& \nonumber \\
&{\displaystyle \exp {\left[ \beta^2 \left( \sum_{\nu >\nu^\prime} q_\nu
q_{\nu^\prime} {\cal G}_{\phi}( t_\nu- t_{\nu^\prime} )+ \sum_{\nu} q_t q_\nu
{\cal G}_\phi(t-t_\nu) +\sum_\nu q_\nu q_0 {\cal G}_\phi(t_\nu) + q_t q_0
{\cal G}_\phi(t) \right) \right]}}&.
\end{eqnarray}
Here $q_\mu= \pm 1$, while $ q_0,q_t = \pm \frac 12$ describe the
contribution of the operators $\psi^\dagger (0)$ and $\psi(0)$, and
${\displaystyle \sum{}^{^{\displaystyle \!\prime}}}$ means that only ''neutral''
configurations with ${\displaystyle \sum_\mu} q_\mu+ q_0 + q_t =0$ are
allowed.  The signs of $q_t$ and $q_0$ depend on a particular
combination of the operators $\psi_{R,L} (0)$ and $\psi^\dagger_{R,L}
(0)$ in Eq. (\ref{eq:fint}). There are altogether four terms. If a
diagonal term $\left< \psi_{R(L)}(0) \ldots \psi^\dagger_{R(L)} (0) \right>
$ is considered, then $q_0 = -q_t = \frac 12$ ($q_0= -q_t=-\frac 12$).
However, if $\left< \psi_{L(R)} (0) \ldots \psi^\dagger_{R(L)} (0) \right>
$ is averaged, then $q_t = q_0 = \frac 12\; (q_t=q_0=-\frac 12$) and here
non-vanishing configurations should contain a compensation charge from
$H^B_{b-sc}$. These compensating charges give rise to factors $e^{\pm
i \varphi_u}$ in the odd power terms of $Z_\phi(t)$.  All the four
combinations have the same dependence on the $\tilde
\phi$-field. Since, as it has been discussed, the
fields~$\phi(x=0,t)$~and~$\tilde \phi(x=0,t^\prime)$ do not interfere,
$F(t)$~factorizes in the form given by Eq.~(\ref{eq:f1}). The fields
$\tilde \phi$ and $\phi$ are related correspondingly to the forward
and the backward scattering channels. The factorization implies a
decoupling of these channels. Another way to prove this decoupling is
a perturbative analysis, in which both the backward and the forward
scattering terms are treated as perturbations. Then the decoupling is
a direct consequence of the vanishing of $\left< \partial_x \phi(x,t)
\phi(y,t^\prime) \right>$ at $x=y=0$. The independence of the forward and
the backward scattering processes has been also realized in a
different way in Refs. \cite{XR:Gogolin93,XR:Kane94}.

The asymptotic behavior of $Z_\phi(t)$ at large time is analyzed by
performing analytical continuation to the Euclidean time $\tau=it > 0$.
Then, $Z_\phi$ acquires a clear physical interpretation --- it becomes a
grand partition function of a classical gas of particles staying on a line
of the ''length'' $\tau$ and interacting via logarithmic Coulomb potentials
$ g(\tau,\tau^\prime)=-qq^\prime \log \left( 1 + \frac{ \alpha v_F
|\tau-\tau^\prime|}{\eta} \right)$. These charged particles should not be
confused with the original electrons; the terms ''length'' and ''distance''
in the discussion of the gas correspond to time intervals in the original
quantum problem. The factor $ z=\frac{\delta_-}{2\pi \alpha}$ is the
fugacity of this gas, while $2\pi/\beta^2=T_{gas}$ acts as an effective
temperature. In addition to the  gas particles two half--charged particles
are fixed to the ends of the system, and only totally  uncharged configurations
contribute.

The physics of such gases has been well studied
\cite{KE&AMM:Anderson70a,RFS:Kosterlitz74b}. There are two
phases separated by the Kosterlitz--Thouless \cite{RFS:Kosterlitz73}
 transition at a critical
temperature $T_{cr}$. At low temperatures, when $T_{gas} < T_{cr}$,
 the particles form dipoles and therefore the interactions of the charges at
the ends are not screened. In the hot phase, when $T_{gas} > T_{cr}$, the
dipoles dissociate and the gas is in the plasma state where logarithmic 
 interactions
between particles are screened--off at distances exceeding the radius of
screening $\tau_{scr}$. The phase diagram of the gas can be obtained by the
renormalization group analysis. It has been shown by Bulgadaev
\cite{EE1D:Bulgadaev82} that in $Z_\phi$, unlike the
systems studied in Refs.
\cite{KE&AMM:Anderson70a,RFS:Kosterlitz74b}, the effective
temperature is not renormalized, while the fugacity scales as
\begin{equation}
\label{eq:rg}
\frac{dz}{d\xi}= z  \left( 1 - \frac {\beta^2}{4\pi} \right),
\end{equation}
where the logarithmic variable $\xi=\log \left( 1+\frac{\tau v_F \alpha}{\eta}
\right)$. For the  dissociation of the dipoles $z$ plays the role of the
activity of the products. If the electron--electron interaction in the
Tomonaga--Luttinger model is attractive ($\gamma<0, \beta^2> 4 \pi$), $z$
scales to zero and, hence, at large distances, there are no free particles
for screening. On the contrary, for the case of repulsion $z$ increases in
the course of renormalization. At a distance
\begin{equation}
\label{eq:tscr}
\tau_{scr} = \frac{\eta}{v_F \alpha} \left( \frac{\delta_-}{s\pi\alpha}
\right)^{-\left( 1-\beta^2/4\pi \right) ^{-1}}
\end{equation}
 the fugacity becomes $\sim 1$, and there are enough particles for
screening.  The numerical constant $s$ can not be determined from
Eq. (\ref{eq:rg}) alone.  We expect it to be of order 1, because the
interpolation procedure discussed at the end of Sec. \ref{se:NL}
indicates that $s \approx 1$.

To analyze the effect of the backward scattering on  the FES, it is
convenient to represent the grand partition function $Z_\phi(\tau)$ as
\begin{equation}
Z_\phi(\tau)= L(\tau) Z(\tau),
\end{equation}
where $Z(\tau)$ is the grand partition function of
the gas without the additional half charges at the
ends, while $L(\tau)$ is the correlation
function of the end charges.  For free electrons $L(\tau)$ is
determined by a series of logarithmically divergent terms. However, the
situation alters entirely for repulsive electron--electron
interaction, because in this case the logarithmic Coulomb interaction
of the charged  particles is screened off.  As it will be shown
bellow the effectively screened Coulomb interaction
decays like a power law of $\tau$ for large enough $\tau$.  The
latter fact allows us to represent the correlation function $L(\tau)$
at $\tau \gg \tau_{scr}$ as
\begin{equation}
\label{eq:Zphi}
L(\tau)= \left( 1+\cos \varphi_u \right)\exp
\left(- \frac{2 \Delta(q= \frac{1}{2})}{T_{gas}} \right)
 \left[ 1+ O(g_{scr}(\tau)/T_{gas}) \right].
\end{equation}
Here $\Delta(q= \frac{1}{2})$ is the self energy of each of the end
charges $q=1/2$ separately, while $g_{scr}(\tau)$ is their effective
mutual interaction; $\Delta(q= \frac{1}{2})$ describes the change of
the free energy of the gas due to the disturbance of the vicinity of
one of the ends, which extends over a distance $\sim \tau_{scr}$.  The
factor 2 appears because the change of the energy of the system due to
the end charges does not depend on the sign of these charges.  The
prefactor $(1+\cos\varphi_u)$ is a result of the factor $\exp\left( i
\varphi_u \sum_\mu q_\mu \right)$ in Eq.~(\ref{eq:fphi}) (see also
Appendix B for a discussion on the effects of $\varphi_u$).

Let us discuss now the screening properties of the gas 
when it is  in the  plasma phase.
 Since the logarithmic interaction is a two dimensional
Coulomb potential, the system can be viewed as a gas of charged
particles confined to a line, while their electric field extends over
the plane. In such a system, despite that a probe charge is
compensated by the charge of the screening cloud, the screening is not
complete, i.e., the potential is not exponentially decaying.
  The equation that determines
the Debye screening of the classical Coulomb gas in the hot plasma
phase is given by
\begin{equation}
\label{eq:Debye}
g_{scr}( y^\prime,y;0,\tau)= \log|y^\prime-y|- \frac{1}{T_{gas}}\int_0^{\tau}
\log|y^\prime- u| \; g_{scr}(u,y;0,\tau) du.
\end{equation}
Here the lengths are measured in units of $\tau_{scr}$, i.e., the scale
above which the classical plasma description can be applied.  The function
$g_{scr}(y^\prime,y;0,\tau)$ describes the potential at a point
$y^{\prime}$ of a probe charge located at $y$ in a system that extends
from $0$ to $\tau$, and it has been used in Eq.~(\ref{eq:Debye}) that
the density of the screening charge $\delta n (u)=
\exp(-g_{scr}(u)/T_{gas})-1 \approx - g_{scr}(u)
/T_{gas}$. When the  probe particle is inside an infinite system one can
easily obtain that the screened potential falls off at large distances as
\begin{mathletters}
\begin{equation}
\label{eq:sins}
g_{scr}(y^\prime,y;-\infty,\infty) \sim \left( \frac{T_{gas} \tau_{scr}}{ y-y^\prime}
 \right)^2 \;\;\;\;\;
\mbox{  for } T_{gas} \tau_{scr} \ll |y-y^\prime|,
\end{equation}
where $|y-y^\prime|$ is the distance from the probe particle
\cite{XR:Decay95}. 
(In this case the probe charge $e^*$ is in the middle of the screening
cloud.  For the one--dimensional geometry this leads to the
configuration of charges: $-1/2e^*,e^*,-1/2e^*$.  This configuration
gives a potential that decays like $1/(y-y^\prime)^2$.) For a finite
system when the probe particle is placed at the end ($y=0$) the
screening is less effective.  For semi--infinite line the
potential  falls off like $(y^\prime)^{-(1+a)}$, where
$0<a<1$. The condition $a>0$ insures the convergence of the integral
over the density of the screening cloud, which is equal to the charge
of the probe particle.  It can be shown (we also checked it
numerically) that
\begin{equation}
\label{eq:send}
g_{scr}(y^\prime,0;0,\infty) \sim  \left( \frac{T_{gas} \tau_{scr}}
{y^\prime} \right)^{1.5}  \;\;\;\;
\mbox{ for } \;\;\; T_{gas} \tau_{scr} \ll  |y^\prime|.
\end{equation}
In the case of a finite interval $(0,\tau)$ there is an ending
effect which leads to a certain increase of  the potential
$g_{scr}(\tau,0;0,\tau) \equiv
g_{scr}(\tau)$. Nevertheless, one can check that the potential
$g_{scr}(\tau)$ still falls off faster than $\tau^{-1}$.  To see that
let us confine the integration on $u$ in Eq.~(\ref{eq:Debye}) to the
point $y^\prime$, neglecting the contribution of particles located
between $y^\prime$ and $\tau$. The effective screening potential
obtained in this way gives an upper bound for the screening potential,
because the contribution of a part of the system to the screening is
ignored.  This procedure reduces the integral equation to the one of Volterra
type, which can be solved by the Laplace transformation.  For $y=0$
this estimate  yields
\begin{equation}
\label{eq:Vol}
g_{scr}(\tau,0;0,\tau) \equiv 
g_{scr}(\tau) < \frac{T_{gas} \tau_{scr}}{\tau \log( \tau/\tau_{scr})}
\;\;\;\; \mbox{for} \;\;\; T_{gas} \tau_{scr} \ll \tau.
\end{equation}
\end{mathletters}
The  fact  that $\tau g_{scr}(\tau)$ decays faster than $\log \tau$ was also
confirmed numerically.

   Since the polarization $\delta n(y^\prime)$ decays rather fast, like
$\left( 1/y^{\prime} \right)^{ 1.5}$, the self--energy of the charge $q=\frac 12$
attached to one of the ends, $\Delta(q=\frac{1}{2})$, is determined by
finite distances $\sim \tau_{scr} T_{gas}$, and it does not contain
any singular contributions.  The effective interaction of the end
charges, $g_{scr}(\tau)$, also does not lead to $\log\tau$ terms in
$L(\tau)$ as it was discussed above.  Therefore, in the hot phase, in
contrast to the marginal non interacting case, there are no
logarithmic terms in $L(\tau)$ and it renormalizes $Z_\phi$ only by a
constant factor.

\begin{figure}[htbp]
\begin{center}
\leavevmode
\epsfbox{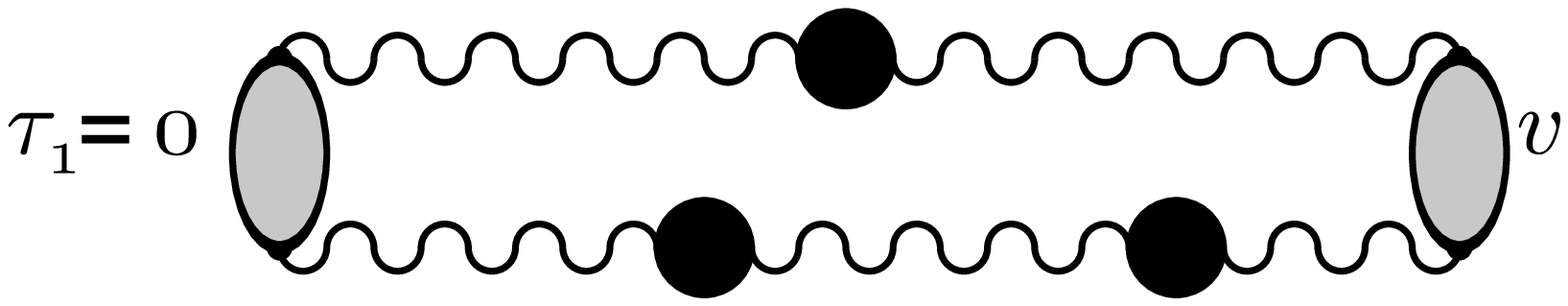}
\end{center}
\caption{ \label{fg:Delta}
A diagram that contributes to $f(\tau)$ after it is integrated with
respect to $v$, which denotes the most distant point to the right. The wavy
lines represent the logarithmic Coulomb interaction between the
charged particles of the gas. The black dots are the polarization
operators of the gas inside the interval $(0,v)$, while the shaded
bubbles are related to the polarization of the gas near the ending
points $0$ and $v$.}
\end{figure}

It is remained now to analyze the contribution of the function
$Z(\tau)$ to the FES.  It is reasonable to write  $Z(\tau)$, which
is the grand partition function of the gas, in the form of the
Mayer--Ursell linked cluster expansion
\begin{equation}
Z(\tau) = e^{C(\tau)}.
\end{equation}
Since $C(\tau)$ is directly related to the thermodynamic potential of
the gas, $C(\tau)=\frac{P \tau}{T_{gas}}$ in the thermodynamic limit
 ($\tau \rightarrow \infty$), where $P$ is the gas pressure.  (When
the analytical continuation back to $t=-i\tau$ is performed, the term
linear in $\tau$ describes the change of the position of the
absorption threshold.  It arises due to a change of the energy of the
electrons in the backscattering potential of the hole.)  To determine
the FES exponent, it is necessary to extract from $C(\tau)$ a singular
sub leading term $\propto \log \tau$, if it exists in the problem
under consideration.  Since the interaction between the charged
particle is effectively short range due to the screening, it is
reasonable to expect that for $\tau \gg \tau_{scr}$ the thermodynamics
of the Coulomb gas behaves in the conventional way. In particular, one
may expect that the effects related to the existence of the two ending
points are not singular in that case, i.e., that sub--leading term $\propto
\log\tau$ does not exist.  To determine 
the sensitivity to the ending points,
let us consider 
\begin{equation}
\label{eq:derC}
f(\tau)= -\left. \frac{\partial C (\tau-\tau_1)}{\partial \tau_1}
\right|_{\tau_1=0}.
\end{equation}
If a $\log(\tau)$ term exists in $C(\tau)$, it should reveal itself in
$f(\tau)$ as $1/\tau$ term.  Taking the derivative in this equation
fixes one of the points in a diagram of the linked cluster expansion
to the left end of the interval $(0,\tau)$.  Let us denote by $v$ the
most distant point in the right direction in each of the diagrams, 
see Fig.~\ref{fg:Delta}.
Then, in each of the diagrams the integration over the coordinates is
limited to the interval $(0,v)$.  Because of the charge neutrality of
the system, the left and right parts of the diagrams can not be
connected by only one line describing the interaction, but should be
connected by at least two interaction lines.  Since the Coulomb 
interaction is long range, these interaction lines should be dressed in
order to reproduce the Debye--H\"uckel screening.  By construction, the
coordinates of the polarization operators in these lines are limited
to the interval $(0,v)$. Therefore, the effective screened potential
corresponding to the interaction lines which connect the left and right parts
of the diagram is $g_{scr}(v,0;0,v)$. The shaded bubbles at the left and
right ends of the diagram in Fig.~\ref{fg:Delta} are equal to a certain 
constant which is determined by short scales $\sim \tau_{scr}$. As a result
\begin{equation}
\label{eq:CON}
f(\tau) \sim \int_0^\tau g_{scr}(v,0;0,v)^2 dv.
\end{equation}
Since $g(v,0;0,v)$ decays faster than $1/v \log v$, a term $\propto
1/\tau$ does not appear in $f(\tau)$.  The diagrams in which the 
left and right ending parts can not be disconnected by cutting two
interaction lines  are not essential, because they decay  with $\tau$ faster 
than the estimate of Eq.~\ref{eq:CON}.
 Therefore, we conclude that $C(\tau)$ does not contain a
logarithmic contribution.

 Finally, after continuing analytically back to real time $t= -i
\tau$, one obtains that in the asymptotic limit 
\begin{equation}
\label{eq:Ztime}
Z_\phi(t) \sim e^{iPt}
\end{equation}
and it  does not contain any power law decaying  preexponential factor.
(This consideration does not exclude the possibility that some terms,
which are exponentially smaller than the main one $ \sim \exp
\left({P\tau/T_{gas}} \right) $, contain a preexponential power law factor.
However, after the analytical continuation such terms will not be at
the Fermi--edge threshold frequency.)

Thus, it has been obtained that in the asymptotic limit, when $\delta \omega =
\omega 
-\omega_{threshold} \ll \tau_{scr}^{-1}$, the absorption line singularity is
given by
\begin{mathletters}
\label{eq:betac}
\begin{equation}
 I(\omega) \propto \left( \delta\omega \right)^{-\alpha_c},
\end{equation}
 where the  exponent of the FES
\begin{equation}
\label{eq:alphac}
\alpha_c=1-\tilde \beta^2/8\pi =
1-\frac{1}{2}
\sqrt{\frac{1+\gamma}{1-\gamma}}+ \frac{\delta_+}{\pi} 
\sqrt{\frac{1-\gamma}{1+\gamma}}  -\frac{1}{2} \left( \frac{\delta_+}{\pi} \right)^2 \left( \sqrt{\frac
{1-\gamma}{1+\gamma}} \right)^3.
\end{equation}
\end{mathletters}
This expression is obtained for the spinless case. In the case of spin 
the result is given by Eq. (\ref{eq:alphas}).

The exponent $\alpha_c$ is non universal and it depends on the interaction
parameter $\gamma$ even when forward scattering is absent ($\delta_+=0$). In
the limit of $\gamma \rightarrow 0$ the FES exponent becomes
\begin{equation} 
\label{eq:alphac0}
\alpha_c(\gamma=0)=\frac 12 + \frac{\delta_+}{\pi} - \frac 12 \left( \frac{\delta_+}{\pi} \right)^2.
\end{equation}
This expression differs from the result of Refs.
\cite{XR:Gogolin93,XR:Prokofev94,XR:Kane94,XR:Affleck94} where
$\frac{3}{8}$ has been obtained instead of $\frac 12$. Although the
difference is relatively small, it arises as a consequence of entirely
different physics at the asymptotic limit, i.e. when $t \gg
\tau_{scr}$.  An interpretation of the above result is presented in the
discussion at Sec. \ref{se:ri} bellow.  The possibility of mapping the
problem onto a Coulomb gas with a characteristic length ($\tau_{scr}$)
indicates a  similarity with the Kondo problem
\cite{KE&AMM:Anderson70a}, where in the asymptotic regime a resonance
singlet is formed. In the Kondo problem the inverse of the
characteristic length is the width of the resonance level at the Fermi
energy, i.e., $ T_K$.  The spin--charge separation in the Kondo
problem is similar to the $\phi$--$\tilde \phi$ channels decoupling
here, while the analogue of the spin singlet (Kondo--resonance) is in
the considered case a ''neutral'' mixture of the left and right moving
electrons.

\section{Non--Local theory}
\label{se:NL}
In the preceding section the analysis of the correlation function
$F(t)$ in the asymptotic limit has been discussed. In this section
$F(t)$ will be studied bellow the asymptotic regime. We
exploit as a starting point the field--theoretical variables, which
has been introduced in Ref. \cite{XR:Kane94}. In these fields the
decoupling of the forward and the backward channels is
transparent. However, at a price of that the problem is described by a
non--local theory of self--dual fields. A special treatment is needed
to handle the theory in the correct way. The renormalization group
analysis of the problem using these variables is elaborated in
Sec. \ref{se:rg}.  Finally a novel iteration procedure, which allows
to find the behavior of $\log F(t)$ at $t < \tau_{scr} $, is
presented in Sec. \ref{se:qtsp}.

\subsection{$\Phi_+ $, $\;\Phi_-$ variables}
\label{se:phi+-}

    The separation of the forward and backward channels can be obtained in a
more transparent way, if instead of the fields $\phi $, $\tilde \phi$
different fields $ \Phi_+, \; \Phi_-$ are employed, which will be defined
now.  In order to introduce these fields one has to construct new operators
of $even$ and $odd$ fermion fields:

\begin{equation}
 \label {eq:idoe}
\begin{array}{c}
\psi_e(k)=\frac{1}{\sqrt{2} } \left( e^{i \varphi_u/2}
\psi_L(-k)+ e^{-i \varphi_u/2} \psi_R( k) \right) \\
\psi_o(k)=\frac{1}{\sqrt{2}i} \left( e^{i \varphi_u/2} \psi_L(-k)- e^{-i
\varphi_u/2} \psi_R( k) \right). \\
\end{array}
\end{equation}
It is easy to check that the operators $\psi_{e (o)}$ have
the standard anticommutation relations. With the use of these fields
the Hamiltonian of noninteracting electrons acquires the form
\begin{equation}
\label{eq:doe}
\begin{array}{c}
H_{free}+H_{sc}=i v_F \int dx
 \Bigl[ 
\psi^\dagger_e(x) \frac{\partial}{\partial x} \psi_{e}(x) + 
\psi^\dagger_o(x) \frac{\partial}{\partial x} \psi_{o}(x) 
\Bigr]+ \\
\int dx \Bigl[ U(0) \Bigl( \rho_e(x) +\rho_o(x) \Bigr) \bbox{\delta}(x) -
|U(2k_F)| \Bigl( \rho_e(x) -\rho_o(x) \Bigr) \bbox{\delta}(x) \Bigr]
 \end{array}
\end{equation} 

Note, that the momenta and the phase dependences in $\psi_{e (o)} (k)$
are adjusted to describe a local--scattering problem $H_{sc}$ in the
most simple representation. Now the bosonization procedure for the
$even$ and the $odd$ fields will be applied. The density
operators will be defined in the usual way
 \begin{equation}
\rho_{e(o)}(q) =\sum_k \psi^\dagger_{e(o)}(k+q) \psi_{e(o)} (k).
   \label{eq:deod}
 \end{equation}
    Both  $\rho_e$-  and  $\rho_o$-operators   have  the  standard  
commutation relations of {\em right} movers:

\begin{equation}
\left[ \rho_{e(o)}(-q),\rho_{e(o)} ( q^{\prime}) \right] = \frac {q L}{2 \pi}
\delta_{q,q^{\prime}}, \;\;\;\;
\left[ \rho_{e}(q),\rho_{o} ( q^{\prime}) \right] =0.
\end{equation}
(This is because in the definition of the operators $\psi_{e(o)}(k)$
the momentum $-k$ is used for the component $\psi_L$.)  Now, the
bosonic representation of the fermion fields can be performed in the
standard way (see e.g. \cite {EE1D:Emery79}):

\begin{equation}
\label{eq:beo}
\psi_{e \left( o \right)}(x)= \frac{1}{ \sqrt { 2 \pi \eta }} 
 e^{i\Phi_{e \left( o \right)} \left( x \right) },
\end{equation}

where
\begin{equation}
  \label{eq:dP}
\Phi_{e (o)} (x) = i \frac{2 \pi}{ L} \sum_p \frac{e^{-ipx - \eta |p|/2}}{p}
\rho_{e (o)} (p)
\end{equation}
     To  complete  the  decoupling procedure  of the forward and
backward scattering processes, two fields are introduced

\begin{equation}
\label{eq:d+-}
\Phi_{\pm}(x) = \Phi_{e} ( x ) \pm \Phi_o ( x ),
\end{equation}
 that satisfy self--dual commutation relations:
\begin{equation}
\label{eq:cPhi}
 \left[  \Phi^\prime_\pm(x), \Phi^\prime_\pm (y) \right]= 4\pi i \partial_y
\bbox{\delta}(x-y); \;\; \left[  \Phi^\prime_+(x),  \Phi^\prime_- (y) \right]= 0
\end{equation}
(here $\Phi^\prime (z) = \frac{d \Phi(z)}{dz}$).
These commutation relations become evident, if one notes that
\begin{equation}
  \Phi^\prime_{e (o)(x)}= 2 \pi \rho_{e (o)} (x).
\end{equation}
Now, the Hamiltonian $H = H_0+H_{sc}$ will be rewritten in terms of
the self--dual fields $\Phi_\pm(x)$. The bosonization of the free
Hamiltonian (\ref{eq:doe}) does not cause any problem, and it remains
to consider the nondiagonal part of the electron--electron interaction
$V \rho_l(x) \rho_R(x)$. From Eqs. (\ref{eq:idoe}) and (\ref{eq:deod})
it follows that
\begin{mathletters}
\begin{equation}
\rho_{L} (q)= \frac{1}{2} \left( \rho_e ( -q )+ \rho_o ( -q ) \right) +
\frac{i}{2} \sum_k \left( \psi^\dagger_e \left( k-q \right) \psi_o \left( k \right) -
                      \psi^\dagger_o \left( k-q \right) \psi_e \left( k \right) \right)
\end{equation}

and

\begin{equation}
\rho_{R} (q)= \frac{1}{2} \left( \rho_e ( q )+ \rho_o ( q ) \right) -
\frac{i}{2} \sum_k \left( \psi^\dagger_e \left( k+q \right) \psi_o \left( k \right) -
                      \psi^\dagger_o \left( k+q \right) \psi_e \left( k \right) \right).
\end{equation}
 Using the bosonic
representation for the last terms one obtains that the Hamiltonian
$H=H_0+H_{sc}$ can be represented as a sum of two commuting parts
\end{mathletters}
\begin{eqnarray}{}
\label{eq:hphi+-}
& H = H_+ + H_- & \nonumber \\
&H_+ = \frac{ v_F}{4 \pi} \int dx \Bigl[ \frac{1}{2 \left( 1 - \gamma \right)} \Phi_+^{\prime}(x)^2 - 2
 \delta_+  \Phi_+^\prime (x) \bbox{\delta}(x)   &
 + \frac {\gamma} {2 \left( 1- \gamma \right)} \Phi_+^\prime ( x ) \Phi_+^\prime ( - x)  \Bigr]\\
&H_- =  \frac{v_F}{4 \pi} \int dx \Bigl[ \frac{1}{2 \left( 1 - \gamma \right)} \Phi_-^{\prime}(x)^2 - 2
 \delta_- \Phi_-^\prime (x) \bbox{\delta}(x)  & +
 \frac{2\gamma}{\eta^2 \left( 1- \gamma \right)}  \left. \cos \Phi_- ( x) \cos \Phi_- (-x) \right)
\Bigr].\nonumber
\end{eqnarray}
A certain care was needed here in order to treat the commutation
relations of the fermion operators properly.
The splitting of $H$ into two independent parts reminds the decoupling
of the fields $\phi(x), \; \tilde \phi(x)$ at a single point (see
Eq. (\ref{eq:0cr})), which was exploited in
Sec. \ref{se:phiphidual}. Now the decoupling of the forward and the
backward scattering channels is performed on a deeper level --- it is
obtained for the Hamiltonian $H = H_+ + H_-$, but not only as a
factorization of the correlation function $F(t)$ in Eq.
(\ref{eq:f1}).

To check the relationship between the fields $\phi(x=0)$, $\tilde
 \phi(x=0)$ and $\Phi_+$, $\Phi_-$ the contribution of the $\Phi_+$
 field to the function $F(t)$ will be calculated.  The Hamiltonian
 $H_+$ is quadratic and  it is possible to perform its diagonalization 
 explicitly.  This can be achieved by the new variables
 \begin{equation}
\label{eq:Theta}
\Theta_+^\prime (x) = \cosh \chi \Phi^\prime_+(x) + \sinh \chi \Phi^\prime_+ (-x),
 \end{equation}
 where $\chi= \frac 12 \mbox{arctanh} \gamma$. Next, the linear
term related with $\delta_+$ can be removed by a canonical
transformation
\begin{equation}
\label{eq:CT}
e^{-iA \Theta_+(y)} \Theta^\prime_+(x) e^{iA \Theta_+(y)} = 
 \Theta^\prime_+(x) - 4 \pi A \bbox{\delta}(x-y),
\end{equation}
(here the self--duality of the fields $\Phi_-$  displays itself --
see for comparison Eq. (\ref{eq:tp})).  Now applying for the
calculation $F(t)$ the bosonic representation (\ref{eq:beo}) one can
verify the fact that $\Phi_+$ channel provides in $F(t)$ exactly the
same factor as the field $\tilde \phi$ in Eqs. (\ref{eq:fb2}) and
(\ref{eq:f1}). Thus, there is a direct relationship between $\tilde
\phi (x=0)$ and $\Phi_+$ and, correspondingly, between $ \phi (x=0)$
and $\Phi_-$.

\subsection{Renormalization of  $\Phi_-$ -- field}
\label{se:rg}

In the variables $\Phi_+$, $\Phi_-$ both the backward and forward
scattering terms are equally simple. That is the main advantage of
this representation. However, the problem under discussion displays a
nontrivial element of the so--called ''dimensional transmutation '',
namely the creation of a new scale -- the screening length
$\tau_{scr}$, which alters the behavior of the system at large scales.
This aspect of the problem is not transparent now. It originates from
the interplay of the backward scattering term and the
electron--electron interaction, while the latter acquires a nontrivial
form in the $\Phi$--fields representation. To demonstrate the relation
of the $\Phi_-$--field theory with the Coulomb gas, the
renormalization group analysis of the Hamiltonian $H_-$ will be
performed. Since the $\Phi_-$--fields posses a nonvanishing
commutation relations, the nonlocality of the interaction term in
$H_-$ demands a certain care. Having this in mind, the interaction term
will be rewritten in the normal ordered form: the components of the
$\Phi_-$ --field with positive momenta (see Eq. (\ref{eq:dP})) should
be placed to the left of the components with negative momenta. The
former are related with the creation operators of the chiral
''phonons'', while the latter with the annihilation operators. As a
result
\begin{equation}
\label{eq:NO}
2 \cos \Phi_-(x) \cos \Phi_-(-x) \Rightarrow f(x) : \cos \Bigl( \Phi_-(x) - 
\Phi_- (-x) \Bigr) :   \;\; ,
\end{equation}
where $f(x) = \frac{1}{1+4 (x/\eta)^2}$ and $: \hat O:$ denotes the
normal ordering of an operator $\hat O$. The $x$-dependent factor
$f(x)$ arises here because the interaction term has a nonlocal and
nonlinear form in fields which do not commute. Only the
derivative--like part in the product of two sines has been kept above,
because the renormalization of the $\delta_-$ term will be studied.

    The transformation from a Hamiltonian to a functional integral
describing quantization of the field with the self--dual commutation
relation (\ref{eq:cPhi}) can be done following Floreanini and Jackiw
\cite{RFT:Floreanini87}.  Finally after passing to the Euclidean space
the effective action is given by
\begin{eqnarray}
\label{eq:p-l}
&S= \int {\cal L} dx d\tau =  
 \frac{1}{4 \pi} \int dx d\tau\Bigl[
-\frac 12 \Bigl( i\dot\Phi_-(x,\tau)+ \Phi_-^{\prime}(x,\tau) \Bigr) \Phi_-^\prime(x,\tau) +& \nonumber \\
&2 \delta_- \left( 1-\gamma
\right) \Phi^\prime_-(x,\tau)\bbox{\delta}(x) - \frac{\gamma}{\eta^2} f(x)
 : \cos \Bigl( \Phi_-(x,\tau)-\Phi_-(-x,\tau) \Bigr) : \Bigr]&,
\end{eqnarray}
where the 'imaginary time' units is rescaled by the factor $v_F/(1-\gamma)$.

    At this stage the regular renormalization group procedure can be
applied.  The variables $\Phi_-(x,\tau)$ in the Lagrangian density will
be divided to fast ($f$) and slow ($s$) components
\cite{RFS:Wilson74}. Here this procedure will be done only with
respect to variation in the $x$--coordinate space, because the local
impurity which is now under consideration is static. In the Fourier
expansion of the slow component the momentum space will be cut off by
 $\tilde \eta = \eta+ d\eta$; see Eq. (\ref{eq:dP}). The renormalization
group transformation will be carried out by progressively integrating
out the fast component and obtaining an effective functional for the
slow component:

\begin{equation}
\label{eq:rg1}
\int {{\cal L}}_s dx d\tau =\log \left( \int {\cal D}\Phi_{-f} \exp \Bigl( {\int {\cal L} dx d\tau} 
 \Bigr) \right)
\end{equation}

where

\begin{equation}
\label{eq:lfs}
{\cal L}={\cal L}^0_s+ \frac{1}{2\pi} \left( 1-\gamma \right) \delta_- \Phi_{-s}^\prime(x,\tau) \bbox{\delta}(x)
   +{\cal L}^0_f+ \frac{1}{2\pi} \left( 1-\gamma \right) \delta_- \Phi_{-f}^\prime(x,\tau) \bbox{\delta}(x)
   +{\cal L}{int},
\end{equation}
and  ${\cal L}^0_{s(f)}=- \frac{1}{8\pi}  \left( i\dot\Phi_{-s(f)}+ \Phi_{-s(f)}^{\prime} \right) 
\Phi_{-s(f)}^\prime $.

To obtain the renormalization of the amplitude $\delta_-$, the
exponent in Eq. (\ref{eq:rg1}) will be expanded in powers of the last
two terms of ${\cal L}$, and the product of these two terms will be
kept. Then, for this product the fast
field $\Phi_{-f}(x,\tau)$ will be integrated out with the weight ${\cal L}_f^0$.
Since the integration of the fast variables leads to a short--range kernel,
it is enough to keep only the quadratic term for the cosine in ${\cal L}_{int}$.
To carry out the integration it is convenient to use the Fourier transform
of the fields $\Phi_{-f(s)}$. As a result one gets 

\begin{equation}
\label{rg:qw}
\Delta S = \frac{2 \delta_- \left( 1-\gamma \right) \gamma}{\left( 2\pi \right) ^3} i \int
d\tilde q d\omega
 \frac{dq}{q} \delta(\omega) \Bigl( f(\tilde q -q) - f(\tilde q +q) \Bigr)
 \Phi_{-s}(\tilde q,\omega),
\end{equation}
 where $\tilde q $ and $q$ denote correspondingly the slow and the  fast
momenta, and $f(k) = \frac{\pi}{2} \eta e^{-|k|\eta/2}$ is the Fourier
transform of the factor $f(x)$. Since $\frac{1}{\eta^2} \Bigl(
f(\tilde q-q)-f(\tilde q +q) \Bigr) \approx \frac{\pi}{2} \tilde q
\mbox{ sign } q $, it follows that
\begin{equation}
\label{rg:qw1}
\Delta S = \frac{2 \delta_- \left( 1 - \gamma \right)}{8 \pi^2} \gamma \xi \int d\tilde q
(i \tilde q) \Phi_{-s}(\tilde q, 0)= \frac{1}{4\pi} \gamma \xi \int dx d\tau
2 \delta_- (1 - \gamma) \Phi_{-s}^\prime(x,\tau) \bbox{\delta}(x)
\end{equation}
where $\xi=log \left( \tilde \eta / \eta \right)$. Exponentiating $\Delta S$
back, one finds that the structure of the $\delta_-$--term has been
reproduced, and the renormalized amplitude $\delta_-( \eta )$ obeys
the equation
\begin{equation}
\label{eq:rgdelta}
\frac{d\delta_-} {d\xi}=\delta_- \gamma,
\end{equation}
which for small $\gamma$ is identical to the renormalization group
 equation of the Coulomb gas theory (see Eq. (\ref{eq:rg})).

\subsection{Canonical transformation}

\label{se:qtsp}

Actually, the renormalization group procedure described above has not
benefited much, in regard to technicalities, by using the
$\Phi_-$--fields. Now, an alternative procedure  which relies on the fact that
in this representation the backward scattering term acquires  a linear
form will be developed.
A canonical transformation, ${\cal U}_A$, generalizing   (\ref{eq:CT}), 
\begin{equation}
\label{eq:CT1}
e^{-i \int A(k) \Phi_-(k) (dk/2\pi)} \Phi_-(p)
e^{ i \int A(k) \Phi_-(k) (dk/2\pi)} =
\Phi_-(p) - i \frac{4 \pi A(p)}{p}
\end{equation}

will be applied
to simplify the Hamiltonian $H_-$, which is taken as:
\begin{equation}
\label{eq:H-}
H_- =  \frac{v_F}{4 \pi} \int dx \Bigl[ \frac{1}{2} \Phi_-^{\prime}(x)^2 - 2
 \delta_- \Phi_-^\prime (x) \bbox{\delta}(x)   +
 \frac{\gamma}{\eta^2}  f(x)  :\cos \Big( \Phi_-(x)- \Phi_-(-x) \Big) :
\Bigr] \; .
\end{equation}
Here the factor $1- \gamma$ has been omitted for brevity, and again
only the derivative--like part of the nonlinear term has been kept.

The aim of the present consideration is to determines the factor $A(k)$
in such a way that as a result of the transformation (\ref{eq:CT1})
the scattering term $\delta_-$ will be removed from the Hamiltonian,
i.e.,
\begin{equation}
\label{eq:trans}
{\cal U}_A^{-1} H_-\left\{ \delta_-\right\} {\cal U}_A = H_-\left\{\delta_-=0\right\} 
\end{equation}
For $A(k)=-\frac{\delta_-}{2 \pi}$ the linear term generated after the
transformation (\ref{eq:CT1}) by the quadratic term in $H_-$ will
cancel out the impurity term. However, because of the nonlinear term,
this is not the end of the story, since
\begin{eqnarray}
\label{eq:cosexp}
&\cos \Bigl( \Phi_- (x) -\Phi_-(-x) \Bigr) \Rightarrow 
\cos \Bigl( \Phi_- (x) -\Phi_-(-x) - 4 \int A(p) \frac{\sin{px}}{p} dp \Bigr) 
\approx&  \nonumber \\
&\cos \Bigl( \Phi_- (x) -\Phi_-(-x) \Bigr) +4 \int A(p) \frac{\sin(px)}{p} dp
\Bigl( \Phi_- (x) -\Phi_-(-x) \Bigr)&.
\end{eqnarray}
Here the cosine was expanded with respect to the $A(p)$--term. In the
Fourier components the new term generated in $H_-$ is given as
\begin{equation}
\label{eq:HA}
H_{sc}^A= -i 8 \frac{\gamma}{\eta^2} \int \frac{dp dk}{ 2\pi} f(p) A(p+k) 
\frac{\Phi_- ( k)}{p+k}. 
\end{equation}
Expanding $f(p)$, one finally gets 
\begin{equation}
\label{eq:HA1}
H_{sc}^A= -2 i \gamma \int dk k \Phi_- (k) \int_0^{\eta^{-1}} \frac{A(p+k)}{p+k}
dp 
\end{equation}
The essential point here is that in this term the combination $k
\Phi_-(k)$ displays the structure of the original $\delta_-$ term.

 The factor $A(k)$ that solves
Eq. (\ref{eq:trans}) can be found by iterations.  For $A(k)= -
\frac{\delta_-}{2 \pi}$ the last integral in Eq. (\ref{eq:HA1}) gives
$\log (1/| k \eta|)$. So, the new $\delta_-$--like scattering term has
been generated with the amplitude $\delta_- \gamma \log ( 1/| k \eta|)$.
Repeating successively the described procedure, one obtains

\begin{equation}
\label{eq:Ak}
A(k)= - \frac{\delta_-}{2 \pi} \Bigl( 1 + \gamma \log ( 1/|k \eta| ) + 
\frac{1}{2} \gamma^2  \log^2 ( 1/ |k \eta| )+ \cdots \Bigr) = - \frac{\delta_-}{2 \pi} e^{\gamma \log ( 1 / |k \eta| )}. 
\end{equation}

We are ready now for the calculation of the correlation function $F(t)$.
Let us write  $F(t)$ as
\begin{equation}
\label{eq:F+-}
F(t)=F^+(t) F^-(t),
\end{equation}
where $F^\pm (t)$ correspond to the correlators of
the fields $\Phi_\pm$ respectively. As it has already been discussed in
Sec. \ref{se:phi+-} the factor $F^+(t)$ can be easily found.  It is
remained to find the correlator 
\begin{equation}
\label{eq:F-}
F^-(t)=
 \left< \exp \Bigl( i H_- \left\{ \delta_- =0 \right\} t \Bigr)
 e^{\frac{i}{2} \Phi_- (0)}
{\cal U}_A  \left[ {\cal U}^{-1}_A 
\exp \Bigl(-i H_- \left\{ \delta_- \right\} t \Bigr)
{\cal U}_A \right] {\cal U}^{-1}_A
e^{-\frac{i}{2} \Phi_- (0)} \right>,
\end{equation}
where ${\cal U}_A$ are determined with $A$ given by Eq. (\ref{eq:Ak}).
It will be assumed here that the essential part of the contribution of
the nonlinear $\gamma$--term has been already taken into account
through the dependence of $A(k)$ on $\gamma$. Therefore, for small
$\gamma$ the averaging over the $\Phi_-$--field in Eq. (\ref{eq:F-}) will
be done keeping in $H_-$ the free term only.  As a result one obtains
\begin{equation}
\label{eq:logft1}
\log F^-(t) \approx \frac{1}{2} \int_0^{\eta^{-1}} dp \Bigl( 1+2A(p) \Bigr)^2
\left[\frac{e^{-iv_Fpt}-1}{p} \right].
\end{equation}

In Sec. \ref{se:phiphidual} the function $F(t)$ was obtained as a
product of two correlators of the fields $\phi$ and $\tilde \phi$.  Since
$F^+(t)$ is equal to the correlator of the $\tilde \phi$--field, it follows
from Eq. (\ref{eq:f1}) that the correlator $F^-(t)$ coincides with $Z_\phi(t)$.
Therefore, from the analysis at  the end of Sec. \ref{se:phiphidual}  one can 
conclude that at asymptotically large time  
\begin{equation}
\label{eq:constf}
F^-(t) \rightarrow const.
\end{equation}
In  the solution (\ref{eq:logft1}) (obtained for 
small enough time) the dependence  on $t$ saturates when 
$1+2A\Bigl( p \sim \frac{1}{v_Ft} \Bigr)$ vanishes, i.e., when  
\begin{equation}
\label{eq:s}
\frac{\delta_-}{\pi} \exp \Bigl( \gamma \log (v_F t / \eta) \Bigr) \approx 1.
\end{equation}
This happens when $t \sim \frac{\eta}{v_F} \left( \frac{\delta_-}{\pi}
\right)^{-\frac{1}{\gamma}}$.
On that  basis it has been  supposed that in Eq. (\ref{eq:tscr})
the parameter $s \approx 1$. 

The described procedure was based on the expansion of the cosine in
Eq. (\ref{eq:cosexp}) after carrying out the canonical transformation
${\cal U}_A$. That expansion is justified when $\left| 4 \int A(p) \frac{\sin
px}{p} dp \right| \lesssim 1 $. For small $\gamma$ it can be shown
that $ 4 \int A(p) \frac{\sin px}{p} dp \simeq 4\pi A(p \sim
\frac{1}{x}) \mbox{ sign } x$.  At the calculation of  $F(t)$ one is  interested
 in momenta
$p \gtrsim \frac{1}{v_f t}$.  Hence, the procedure is safe until
 $\left| 4 \pi A(p \sim \frac{1}{v_F t}) \right| \lesssim 1$, i.e., when
\begin{equation}
\label{eq:con}
2{\delta_-}\exp \Bigl( \gamma \log (v_F t / \eta) \Bigr) \lesssim 1.
\end{equation}
Thus, $F^-(t)$ has been found for asymptotically large time $ t \gg
\tau_{scr}$ when the physics of the screened Coulomb gas develops, and
also for short time when $\gamma \log \Bigl( \frac{v_F t} {\eta}
\Bigr) \lesssim \log \Bigl( \frac{1 }{2\delta_-} \Bigr)$.  Therefore,
the point determined above from Eq. (\ref{eq:logft1}) where $F^-(t)$
saturates is, in fact, only an interpolation estimate: the behavior of
$F^-(t)$ has been found in two limits, and the two solutions are
matched at a time which should determine $\tau_{scr}$.

\section{Discussion: Resonance interpretation}
\label{se:ri}
 To interpret the obtained result for the  exponent $\alpha_c$ one
can try to apply the Nozi\'eres and De Dominicis \cite{XR:Nozieres69c} (ND)
 theory of the FES for one dimension. Then a serious
problem arises, because the ND theory uses the scattering theory
description. This implies the existence of quasi-particles. However, as it
is known from studies of the Tomonaga--Luttinger model
\cite{EE1D:Dzyaloshinskii74,EE1D:Luther74}, quasi-particles do not exist in
one dimension when $\gamma$ is finite. Therefore, it has sense to discuss
the physical meaning of $\alpha_c$ in terms of the scattering theory
description only in the limit of vanishing $\gamma$, i.e. for Eq.
(\ref{eq:alphac0}). At this limit the screening length $\tau_{scr}(\gamma)$
goes to infinity, and the following discussion corresponds to
\begin{equation}
\label{eq:limits}
t \gg \tau_{scr} \rightarrow \infty.
\end{equation}

 In the spinless case the ND theory yields $\alpha_c= 2 \frac {\delta_{l_0}}
{\pi} - \sum_l (2l+1) \left( \frac{\delta_l}{\pi} \right)^2$, where $\delta_{l}$ is
the phase shift of the spherical harmonic component $l$ and $\delta_{l_0}$
is the phase shift of the state of the exited electron. In the case of one
dimension the even and odd combinations replace the partial wave expansion
\cite{RFS:Lipkin73}. Hence in the one dimensional case the ND theory gives
\begin{equation}
\label{eq:FESep}
 \alpha_c= 2 \frac{\delta_e}{\pi}-\left( \frac{\delta_o}{\pi} \right) ^2- \left(
\frac{\delta_e}{\pi} \right) ^2
\end{equation}
when the even state is excited (see Appendix \ref{ap:odd} for a 
discussion on the odd state).
Here $e^{2i\delta_{e (o)}}$ are the eigenvalues of the $2 \times 2$ unitary
scattering matrix of the one--dimensional system. 
 The relation of the phase shifts with the transmission  ($t_s$) 
and the reflection ($r_s$) scattering amplitudes can be easily found:
\begin{equation}
\label{eq:selfphases}
\delta_{e (o)}= \frac 12 \left( \varphi_t \pm \arctan \frac{|r_s|}{|t_s|} \right),
\end{equation}
where $\varphi_t$ is the argument of the amplitude $t_s=|t_s|e^{i\varphi_t}$.

To determine the phase shifts for the asymptotic limit (\ref{eq:limits}) 
let us discuss 
the scattering of particles with a linearized spectrum, the appropriate
 Schr\"odinger equation is 
\begin{equation}
\label{eq:22dif}
\left( \begin{array}{cc}  iv_F \frac{d}{dx} + U(k=0)\bbox{\delta}(x) & U(2k_F) \bbox{\delta}(x) \\
             U^*(2k_F) \bbox{\delta}(x) & -iv_F \frac{d}{dx} + U(k=0)\bbox{\delta}(x)\\
   \end{array} \right) \left( \begin{array}{c} \Psi_R(x) \\ \Psi_L(x) \end{array} \right) =
  E \left( \begin{array}{c} \Psi_R(x) \\ \Psi_L(x) \end{array} \right).
\end{equation}
The eigenfunctions of that equation can be easily found
\begin{equation}
\label{eq:ef}
\left( \begin{array}{c} \Psi_R(x) \\ \Psi_L(x) \end{array}\right)_{e(o)} \propto
\left( \begin{array}{c}
     e^{-i\left( \frac{E}{v_F}x +\frac 12(\delta_+ \pm \delta_-)\mbox{ sign }(x)
-\frac{\varphi_u}{2} \right)}\\
 \pm e^{ i\left( \frac{E}{v_F}x +\frac 12(\delta_+ \pm \delta_-)\mbox{ sign }(x)
-\frac{\varphi_u}{2} \right)}
\end{array} \right)
\end{equation}
where $\delta_+=-U(k=0)/v_F$, $\delta_-=|U(2k_F)|/v_F$ and 
$ \varphi_u = \arg \left( -U(2k_F) \right)$. The phase shifts of this solution 
 are given by
\begin{equation}
\label{eq:ephases}
\delta_{e (o)}= \frac 12 \left( \delta_+ \pm \delta_- \right).
\end{equation}
(Here the decoupling of  the forward and backward channels reveals itself in
 the fact that $\delta_e \pm \delta_o = \delta_{\pm}$. It should be
mentioned, however, that the use of the operator $iv_F \frac{\partial}{\partial
x}$ for the linearized spectrum of the electrons neglects at the description
of the forward scattering the presence of the cutoff in the momentum space.
When one solves Eq. (\ref{eq:22dif}) by the standard scattering formalism
 with a finite cutoff in the spectrum the decoupling does not occur.
 Nevertheless, it is reasonable to follow the present treatment of the 
 scattering  since from the
experience of studies  of the $X$-ray absorption problem
\cite{XR:Nozieres69c,XR:Schotte69a}  and
the Kondo problem \cite{KE&AMM:Yuval70,KE&AMM:Anderson70a,KE&AMM:Schotte70} 
it is  known that the bosonization procedure  provides the correct mapping
 on  the Coulomb
gas if one replaces the Born phase shift by the full one.)
Substitution of the phase shifts (\ref{eq:ephases}) into Eq. (\ref{eq:FESep}) and comparison
with Eq. (\ref{eq:alphac0}) yield
\begin{equation}
\label{eq:resonancephase}
\delta_{e(o)}= \frac{\delta_+}{2} \pm \frac{\pi}{2}.
\end{equation}
From the general formalism  of the scattering theory it is known that when a
resonance exists in a channel the phase shift of this channel is close to
$\pm \pi/2$ .

The result  of  Refs.
\cite{XR:Gogolin93,XR:Prokofev94,XR:Kane94,XR:Affleck94} was based on a 
different assumption.
 Comparison of  Eqs. (\ref{eq:ephases}) and (\ref{eq:selfphases})
gives
\begin{equation}
\label{eq:rdelta}
\varphi_t=\delta_+; \;\;\;\;\;\;\; \arctan(\frac{|r_s|}{|t_s|})=\delta_-.
\end{equation}
 In the course of the renormalization the strength of the
backward scattering amplitude increases. Therefore, in  view of Eq. (\ref{eq:rdelta}) it is
tempting to accept  that the asymptotic limit corresponds to the  total reflection.
Then, according to Eq. (\ref{eq:rdelta}), $\delta_-= \pi/2$, and  
 from Eqs. (\ref{eq:FESep}, \ref{eq:ephases}) it can be obtained 
$\alpha_c=\frac{3}{8}+\frac{\delta_+}{\pi}-\frac 12\left( \frac{\delta_+}{\pi} \right)
^2 $. Precisely that result was given  in Refs.
\cite{XR:Gogolin93,XR:Prokofev94,XR:Kane94,XR:Affleck94}.
In fact, it has been assumed in Ref. \cite{XR:Affleck94} 
 as a starting point that the reflection
coefficient is one, and  the appropriate boundary condition  has been
applied. The same assumption has been also employed  in Ref.
\cite{XR:Prokofev94}.

Indeed, following Eq. (\ref{eq:rdelta}) it is difficult to imagine how the 
limit of a total reflection as the asymptotic one could be escaped.
The answer was formulated in the first paragraph of this discussion. 
The point is that  the interpretation in terms of the scattering phases
can be applied  only in the asymptotic limit of free particles.
 At the intermediate scales this simple  interpretation
 is not valid.
  The same effects of electron--electron interaction which lead to the
renormalization of the backward scattering make the use of the theory of
scattering not applicable at the intermediate scale. Therefore, one cannot
apply Eq. (\ref{eq:rdelta}) during the course of the renormalization procedure.
Consequently, the argument that one can not get a value of the FES
exponent larger than $3/8$ without crossing the regime corresponding
to the total reflection can not be used  against the result for 
$\alpha_c$ obtained in Eq.~(\ref{eq:alphac}).

Another point, which is worth to be discussed, is the intimate
relation of the result of
Refs. \cite{XR:Gogolin93,XR:Prokofev94,XR:Kane94,XR:Affleck94} with
the work of Kane and Fisher \cite{EE1D:Kane92}.  In the latter it has
been claimed that, since in the course of renormalization flow the
strength of the impurity effectively increases, the final fixed point
can be modeled by two disconnected semi--infinite lines (wires). That kind of
fixed point leads to a total reflection and is in accordance with the
results of Refs.
\cite{XR:Gogolin93,XR:Prokofev94,XR:Kane94,XR:Affleck94} for the FES.
 However, two weakly connected
wires represent a system with severely broken particle--hole symmetry  at the
impurity site because   electron density vanish at the ends of the wires.
 To break  the particle--hole 
symmetry a depletion of electrons must occur at the impurity area.
  The total depleted charge, $\delta {\cal N}$, according to Friedel sum rule
 \cite{RFS:Friedel52,RFS:Friedel58,RFS:Langer61}, is equal to the sum of the phase shifts, i.e. according to Eq.
 (\ref{eq:ephases})
\begin{equation}
\label{eq:Friedel}
\delta {\cal N} = \frac{\delta_e+\delta_o}{\pi}=\frac{\delta_+}{\pi}. 
\end{equation}
Therefore the term responsible for the particle--hole asymmetry is the
amplitude $U(k=0)$, which in contrast to $U(2k_F)$ does not increase
in the course of the renormalization flow.  Thus, two disconnected
semi--infinite lines cannot be a fixed point to which a weak impurity
system flows.

It has been attempted \cite{EE1D:Kane92a} to justify the conjecture of
two disconnected semi--infinite lines as the fixed point limit by
mapping the problem for a particular value of the interaction coupling
constant on an exactly solvable model of a semi--infinite spin chain
\cite{EE1D:Guinea85}.  To make this mapping onto a semi-infinite
geometry possible, the authors had to impose an additional constraint
on the field $\tilde \phi(x=0,\tau)$.  It was needed to set $\tilde
\phi(x=0,\tau)=0$ (see the discussion of Eq. (8.3b) in
Ref. \cite{EE1D:Kane92a}), what implies that the charge fluctuations
are frozen out at the location of the impurity center. Indeed, this
situation can be realized in a weak--link junction.  However, the
vanishing of the charge density fluctuations is, in fact, not a result
of the dynamics of the original problem, but is a direct consequence
of the additionally imposed constraint.

It has been shown that the asymptotic behavior of the 1--d electron
backward scattering in case of electron--electron repulsion resembles
the physics of the Kondo resonance.  A similarity with the Kondo
problem is seen in the possibility of mapping the considered above
problem onto a Coulomb gas with a characteristic length $\tau_{scr}$.
In the considered problem a localized mixture of the left and right
moving electrons acts the role of the Kondo singlet.  In Eq. (44) of
Ref. \cite{EE1D:Matveev95} a Hamiltonian quadratic in fermion
operators is presented, which is similar to the Hamiltonian of a
resonant--level model. The partition function of this model
corresponds to a non alternating Coulomb gas with $\beta = \sqrt{2
\pi}$.  It is also claimed in Ref. \cite{EE1D:Matveev95} that at low
energies this resonant--level Hamiltonian is equivalent to the one
describing the two--channel Kondo problem in the Toulouse
limit\cite{KE&AMM:Emery92}.  Thus, the FES problem in interacting
one--dimensional electron gas analyzed above by mapping on a non
alternating Coulomb gas has indeed a certain relationship with a Kondo
resonance physics.
However, it should be emphasized that the interpretation of the obtained
result for the FES in terms of the phase shifts should not be taken
too literally. 
At the consideration of other physical quantities the phase shifts of
Eq.~(\ref{eq:resonancephase}) can not be used straightforwardly,
because in one--deimension an electron gas with interactions cannot be described
by the Fermi--liquid theory.

\section{Conclusions}
\label{se:Conclusions}

The absorption of the electro-magnetic wave in one-dimensional
electron systems has been analyzed near the absorption edge.  It has
been found that a new time scale, $\tau_{scr}$, is generated as a
result of the combined effect of the backward scattering of electrons
on an impurity--like center created in the valence band at the
absorption and a repulsive interaction of the conduction electrons.
The shape of the absorption line is given by the Fourier transform of
the correlation function $F(t)$, the long time asymptotic behavior of
which is determined by $\tau_{scr}$.  Consequently for frequencies
close enough to the Fermi--edge the absorption line is controlled by
this scale. The infrared physics of the Fermi--edge singularity in the
presence of backward scattering together with electron--electron
repulsion resembles the physics of the Kondo problem. This aspect of
the Fermi--edge singularity in 1--d systems was missed in the
preceding studies of the question.  The approach of the present paper
may not be confined to the Fermi--edge singularity problem only. It
may be useful in the studies of tunneling effects in  quantum wires and
 also for description of effects related to the quantum Hall edge states.

\acknowledgments
We are grateful to G.~Kotliar, A.~Kamenev and D.~Orgad for useful and
stimulating discussions. A.~F. is supported by the Barecha Fund Award.
The work is supported by the Israel Academy of Science under the Grant
No. 801/94-1.

 \appendix
\section{One--Dimensional FES for Spin Case}
\label{app}

When the spin degrees of freedom of the conduction electrons are
included, the main features of the theory of the FES in 1--d do not
change.  The forward scattering contribution can be found by a
shifting operator, while the backward scattering contribution can be
described by a reduction to the Coulomb gas theory with a
characteristic screening length. Due to the spin, the charged plasma
contains two types of particles.  The fields that describe the forward
and backward channels remain to be decoupled. The final expression for
the FES exponent have the same structure as in the case of spinless electrons,
but with  slightly different coefficients. 

The electron--electron interaction in the presence of spin is taken as 
\begin{equation}
\label{eq:H0spin}
H^s_{int}= \frac 12 V\int \left( \rho_{R\uparrow}(x) + \rho_{R\downarrow}(x) + \rho_{L\uparrow}(x) +
\rho_{L\downarrow}(x) \right)^2 dx,
\end{equation}
where $\rho_{i\mu}(x)=\psi^\dagger_{i \mu}(x) \psi_{i \mu}(x); \;
i=R,L$ and $ \mu=\uparrow,\downarrow$.
The forward and backward scattering terms are given by

\begin{equation}
\label{eq:Hsfw}
H^s_{f-sc}=U(k=0) \left[ \rho_{R \uparrow}(0)+ \rho_{R \downarrow}(0)+ \rho_{L \uparrow}(0) + \rho_{L
\downarrow}(0) \right]
\end{equation}
and 
\begin{equation}
\label{eq:Hsbw}
H^s_{b-sc} =  U(2k_F) \left( \psi^\dagger_{R \uparrow} (0) \psi_{L \uparrow} (0) + \psi^\dagger_{R
\downarrow} (0) \psi_{L \downarrow}(0)\right) + h.c.
\end{equation}
Now the bosonization representation for up and down spins will be applied in a way 
similar to the spinless case (see Eqs. (\ref{eq:dpd}) and  (\ref{eq:fermion})).
With the use of the conventional combinations $\rho$ and $\sigma$

\begin{mathletters}
\label{def:psdps}
\begin{eqnarray}
\phi_\rho=\frac{1}{\sqrt 2} \left( \phi_\uparrow + \phi_\downarrow \right) ; \;
\phi_\sigma=\frac{1}{\sqrt 2} \left( \phi_\uparrow - \phi_\downarrow  \right) \\
\tilde \phi_\rho=\frac{1}{\sqrt 2} \left( \tilde \phi_\uparrow + \tilde \phi_\downarrow \right); \;
\tilde \phi_\sigma=\frac{1}{\sqrt 2} \left( \tilde \phi_\uparrow - \tilde \phi_\downarrow \right)
\end{eqnarray}
\end{mathletters}
 one arrives to the bosonized Hamiltonian $H^s=H_0^s+H^s_{f-sc}+H^s_{b-sc}$:

\begin{eqnarray}
\label{eq:Hsp}
&H^s_0= \frac 12 v_F \alpha_\rho \int dx \left( \left( \frac{d\tilde \phi_\rho  }{dx} \right)^2    +
                                          \left( \frac{d       \phi_\rho  }{dx} \right)^2 \right) +
         \frac 12 v_F             \int dx \left( \left( \frac{d\tilde \phi_\sigma}{dx} \right)^2    + 
                                          \left( \frac{d       \phi_\sigma}{dx} \right)^2 \right), & \\
&H^s_{f-sc}=  \frac{v_F \delta_+}{\sqrt{2} \pi} \beta_\rho\frac{d\phi_\rho}{dx},  &   \\
&H^s_{b-sc}= -\frac{v_F \delta_-}{\eta \pi}
               \Bigl[ \cos \Bigl( \sqrt{4\pi} \phi_\uparrow(0) + \varphi_u \Bigr) +
                      \cos \Bigl( \sqrt{4\pi} \phi_\downarrow(0) + \varphi_u \Bigr) \Bigr],&
\end{eqnarray}
where $\alpha_\rho= \frac{4\pi}{\beta_\rho^2}$, $\beta_\rho^2=4\pi \sqrt {
\frac{1-\gamma_\rho}{1+\gamma_\rho}}$ and  $\gamma_\rho=\frac{V}{\pi v_F + V}
=\frac{2 \gamma}{1+\gamma}$.

In the backward scattering term the $\rho$ and $\sigma$ channels
appear to be mixed.  In the absence of the backward scattering the
FES can be found by shifting $\frac{d\phi_\rho}{dx}$ with the use of
the exponent of the dual operator $\tilde \phi_\rho$. The forward
scattering appears only in the $\rho$ channel, and in comparison to
the spinless case it acquires an additional factor of $\sqrt{2}$.
Finally, the expression for the FES exponent when spin is included,
but only  the forward scattering exists. is
\begin{equation}
\label{eq:betas}
\alpha^s_{forward}= \frac 12 \left( 1- \frac{1}{\sqrt {1-\gamma_\rho^2} } \right)
 +\frac{\delta_+}{\pi}\sqrt{\frac{1-\gamma_\rho}{1+\gamma_\rho}}
 -\left( \frac{\delta_+}{\pi} \right)^2 \left( \sqrt{\frac{1-\gamma_\rho}{1+\gamma_\rho}} \right)^3.
\end{equation}
This result coincides with the result of Ogawa
et. al. \cite{XR:Ogawa92}, if in the corresponding expression of Ref.
\cite{XR:Ogawa92} one substitutes $v_F$ by $\frac{v_F}{1-\gamma}$,
$g_2$ by $v_F \frac{\gamma}{1-\gamma}$ and $g^{cv}$ by $2
\frac{\delta_+}{\pi} v_F$. The renormalization corrections arise,
because the interaction between the electrons moving in the same
direction have been included here.

The correlation function $F(t)$ in the presence of the backward
scattering will be treated analogously to Sec. \ref{se:bs}. One should
calculate an expression equivalent to Formula (\ref{eq:fint}), but for
the discussed case it has a slightly more complicated form:
\begin{eqnarray}
\label{eq:Fstt}
F(t)= &{\displaystyle \sum_{p,q=\pm}} \left<
 e^{-i  \frac 12 \left( \sqrt {4\pi} \tilde \phi^0_\uparrow(t) + p\sqrt{4 \pi}
\phi^0_\uparrow(t) \right) } 
e^{i \frac{1}{\sqrt{2}} \frac{\delta_+}{\pi}
\frac{\beta_\rho}{\alpha_\rho}\tilde \phi_\rho(t)} \right.& \Bigl[ \cdots \nonumber\\
\cdots \Bigr] &\left.
e^{-i \frac{1}{\sqrt{2}} \frac{\delta_+}{\pi}
\frac{\beta_\rho}{\alpha_\rho}\tilde \phi_\rho(0)}
e^{i  \frac 12 \left( \sqrt {4\pi} \tilde \phi^0_\uparrow(0) + q\sqrt{4 \pi} \phi^0_\uparrow(0)
\right)}
\right>.& 
\end{eqnarray}
Here the dots inside the square brackets denote the time integrals  of  
a sum of all possible time--ordered products of the exponentials of the 
operators $\pm i \left( \sqrt{4\pi} \phi^0_{\uparrow(\downarrow)}+ \varphi_u \right) $.
Only those terms which correspond to equal number of creation 
and annihilation  operators of electrons, separately for each spin species, 
give non vanishing contributions. 

Since the fields $\phi(x=0,t)$ and $\tilde \phi(x=0,t^\prime)$ do
 not interfere, $F(t)$ is factorized as 
\begin{equation}
\label{eq:spin-factor}
F(t)=e^{ -\frac{1}{8} \left( \tilde \beta_\rho^2 {\cal G}_\rho (t) 
         + 4 \pi  {\cal G}_\sigma (t) \right) }
                          Z_s(t),
\end{equation}
where $\tilde \beta_\rho = \frac {4 \pi}{\beta_\rho}- 2 \frac{\delta_+
 \beta_\rho}{\pi \alpha_\rho}$, ${\cal G}_{\rho}(t)= \left<
 \phi_{\rho}(0,t) \phi_{\rho}(0,0)-\phi_{\rho}(0,0)^2 \right>= \frac{1}{2
 \pi} \log(1+ i \alpha_{\rho} v_F t/\eta)$ and 
${\cal G}_{\sigma}(t)= \left<
 \phi_{\sigma}(0,t) \phi_{\sigma}(0,0)-\phi_{\rho}(0,0)^2 \right>= \frac{1}{2
 \pi} \log(1+ i  v_Ft/\eta)$. 
The correlator $Z_s(t)$ can be obtained similarly to Eq. (\ref{eq:fphi}):
\begin{eqnarray}
\label{eq:fphis}
&{\displaystyle Z_{s}(t)= \sum_{n=0}^\infty \left( \frac{ \delta_-}{2\pi
} \right) ^n
  \sum_{q_t,q_0,q_{\mu},\lambda_0} {}^{^{\displaystyle \!\!\!\!\prime}}
 i^n  e^{\left( i \varphi_u \sum_\mu q_\mu \right)} 
\int_0^t \frac{ v_F}{\eta} dt_n \ldots \int_0^{t_3} \frac{
v_F}{\eta} dt_2
\int_0^{t_2} \frac{ v_F}{\eta} dt_1}& \nonumber \\
&{\displaystyle \exp {\left[  \left( \sum_{\nu >\nu^\prime, \lambda, \lambda^\prime} q_\nu
q_{\nu^\prime} {\cal D}_{\lambda \lambda^\prime}( t_\nu- t_{\nu^\prime} )+
 \sum_{\nu, \lambda} q_t q_\nu
{\cal D}_{\lambda_0 \lambda} (t-t_\nu) +\sum_{\nu, \lambda} q_\nu q_0
 {\cal D}_{\lambda \lambda_0}(t_\nu) + q_t q_0
{\cal D}_{\lambda_0 \lambda_0}(t) \right) \right]}}&.
\end{eqnarray}
Here indices $\lambda$ denote the spin projections $\uparrow, \downarrow$;
${\cal D}_{\uparrow \uparrow}={\cal D}_{\downarrow \downarrow}= \frac 12 \left( \beta_\rho^2 {\cal G}_\rho + 4\pi {\cal G}_\sigma \right)$
and
${\cal D}_{\uparrow \downarrow}={\cal D}_{\downarrow \uparrow}= \frac 12 \left( \beta_\rho^2 {\cal G}_\rho - 4\pi {\cal G}_\sigma \right)$.
The sum $Z_s(t)$ repeats the structure of $Z_\phi(t)$ in
Eq. (\ref{eq:fphi}) with the only difference that now there are two
sorts of Green functions. Correspondingly ${\displaystyle \sum
{}^{^{\displaystyle \!\prime}}}$ means that only ''neutral'' configurations for each
spin species separately are allowed.  Now the discussion following
Eq. (\ref{eq:fphi}) in Sec. \ref{se:bs} that leads to a Coulomb gas
theory can be repeated. In the case of spin this gas in addition to
charges has two flavors. Charges of the same flavor interact through
$D_{\uparrow \uparrow}$, while charges with different flavor interact through
$D_{\uparrow \downarrow}$.
The screening properties of the plasma of such gas are similar to those
of the single component case. As a result, in  the asymptotic limit
when $\delta \omega = \omega - \omega_{treshold} \ll \tau_{scr}^{-1}$
the exponent describing the absorption  singularity is 
given by
\begin{equation}
\label{eq:alphas}
\alpha_s=  \frac{3}{4} - \frac{1}{4} \sqrt{\frac{1+\gamma_\rho}{1-\gamma_\rho}}
 +\frac{\delta_+}{\pi}\sqrt{\frac{1-\gamma_\rho}{1+\gamma_\rho}}
 -\left( \frac{\delta_+}{\pi} \right)^2 \left( \sqrt{\frac{1-\gamma_\rho}{1+\gamma_\rho}}
\right)^3.
\end{equation}

\section{Excitation of Even and Odd modes}
\label{ap:odd}
Let us define creation operators of the even and odd eigenmodes according to  
Eq. (\ref{eq:ef})

\begin{equation}
\psi^\dagger_{e} = \frac{1}{\sqrt{2}}
 \left( e^{ i \frac{\varphi_u}{2}} \psi^\dagger_R + 
                e^{-i \frac{\varphi_u}{2}} \psi^\dagger_L \right) \; , \;
\psi^\dagger_{o} = \frac{1}{\sqrt{2}i}
 \left( e^{ i \frac{\varphi_u}{2}} \psi^\dagger_R -
                e^{-i \frac{\varphi_u}{2}} \psi^\dagger_L \right).
\end{equation}
Now let us consider $Z_\phi^{e(o)}(t)$ corresponding to activation of
the states $\psi^\dagger_{e(o)}$ at the light absorption, instead of
the combination $\psi^\dagger_L+ \psi^\dagger_R$ which has been
studied in Eq.  (\ref{eq:fint}).  For all configurations which give
non vanishing contributions to $Z_\phi$ the factors depending on the
phase $\varphi_u$ are canceled out when the eigenmodes operators
$\psi^\dagger_{e(o)}$ are used.  Then $Z_\phi^{e}(t)$ corresponds to a
maximum, while $Z_\phi^{o}(t)$ to a minimum, of the function
$Z_{\phi}(t)$, when one continues it analytically to the Euclidean
time $\tau= it >0$ and studies $Z_\phi$ as a function of
$\varphi_u$. This is because in the Coulomb gas expansion of
$Z_\phi^e$ all terms become positive, while in the $Z_\phi^o$ sum the
terms corresponding to configurations with odd number of gas particles
have opposite sign. Therefore, the FES corresponding to the excitation
of the odd channel is less singular.
This is in accordance with  Eq. (\ref{eq:Zphi}),
 because the odd combination corresponds to $\varphi_u=\pi$ when the factor
$1+\cos \varphi_u$ vanishes.


\end{document}